\documentclass[aps,reprint,graphicx]{revtex4-1}
\usepackage{amsmath, graphicx}
\usepackage{xcolor}
\usepackage{hyperref}
\hypersetup{colorlinks=true,allcolors=black}

\begin{document}

\title{Non-Markovian effects in long-range polariton-mediated energy transfer} %Title of paper

\author{Kristin B Arnardottir}
\email[]{arnardottir@mci.sdu.dk}
%\homepage[]{Your web page}
%\thanks{}
%\altaffiliation{}
\affiliation{POLIMA---Center for Polariton-driven Light--Matter Interactions, University of Southern Denmark, Campusvej 55, DK-5230 Odense M, Denmark}
\affiliation{SUPA, School of Physics and Astronomy, University of St Andrews, St Andrews,
KY16 9SS, UK}
\author{Piper Fowler-Wright}
\affiliation{SUPA, School of Physics and Astronomy, University of St Andrews, St Andrews,
KY16 9SS, UK}
\affiliation{Department of Chemistry and Biochemistry, University of California San Diego, La Jolla, CA 92093, USA}
\author{Christos Tserkezis}
\affiliation{POLIMA---Center for Polariton-driven Light--Matter Interactions, University of Southern Denmark, Campusvej 55, DK-5230 Odense M, Denmark}
\author{Brendon W Lovett}
\affiliation{SUPA, School of Physics and Astronomy, University of St Andrews, St Andrews, KY16 9SS, UK}
\author{Jonathan Keeling}
\affiliation{SUPA, School of Physics and Astronomy, University of St Andrews, St Andrews, KY16 9SS, UK}

\date{\today}

\begin{abstract}
Intramolecular energy transfer driven by near-field effects plays an important role in applications ranging from biophysics and chemistry to nano-optics and quantum
communications. Advances in strong light--matter coupling in molecular systems have opened new possibilities to control energy transfer. In particular, long-distance energy transfer between molecules has been reported as the result of their mutual coupling to cavity photon modes, and the formation of hybrid polariton states. In addition to strong coupling to light, molecular systems also show strong interactions between electronic and vibrational modes. The latter can act as a reservoir for energy to facilitate off-resonant transitions, and thus energy relaxation between polaritonic states at different energies. However, the non-Markovian nature of those modes makes it challenging to accurately simulate these effects. Here we capture them via process tensor matrix product operator (PT-MPO) methods, to describe exactly the vibrational environment of the molecules combined with a mean-field treatment of the light--matter interaction. In particular, we study the emission dynamics of a system consisting of two spatially separated layers of different species of molecules coupled to a common photon mode, and show that the strength of coupling to the vibrational bath plays a crucial role in governing the dynamics of the energy of the emitted light; at strong vibrational coupling this dynamics shows strongly non-Markovian effects, eventually leading to polaron formation. Our results shed light on polaritonic long-range energy transfer, and provide further understanding of the role of vibrational modes of relevance to the growing field of molecular polaritonics. 
\end{abstract}


\maketitle

\section{Introduction}

Resonant energy transfer between molecules
plays a crucial role in biochemical and photophysical
processes, where excited donor molecules transfer their energy
to acceptor molecules through nonradiative dipole
coupling~\cite{scholes_arpc54,vangrondelle_pccp8}. 
F\"{o}rster resonant energy transfer is typically a short-range
effect, based on the near-field interaction between two dipoles,
and with rates of transfer scaling approximately as $d^{-6}$ where $d$ is the
dipole separation~\cite{forster_annphys437}.
However, the enery transfer process can be
modified by tailoring the local density of states at the
donor emission frequency, e.g.\ by placing the molecules
in an optical cavity~\cite{andrew_sci290}. In recent years,
following the rapid growth of the field of molecular polaritonics~\cite{torma_rpp78,keeling20,basov_nanoph10}, there has
been a series of explorations of energy transfer between two
spatially separated excitonic modes mediated through a common
strongly coupled photon mode\cite{saez_prb97,saez_aom8,georgiou_ultralong-range_2021,pajunpaa_polariton-assisted_2024,hu_energy_2024}.
The ability to control such transfer processes by tuning light--matter
interactions has a range of applications, such as energy harvesting and quantum communications~\cite{chanyawadee_prl102,vogelsang_natmat10,blum_prl109}.

As the separation between molecules in polaritonic
cavities is typically much larger than the F\"{o}rster
radius, an alternative mechanism must be involved to explain
such experiments on polariton-mediated energy transfer. 
Collective strong coupling to light combined with local
coupling to vibrational baths has been suggested as the
governing mechanism in these situations; such dynamics has been
explored within Redfield theory~\cite{saez_prb97}. In the
presence of a mediating cavity, the coupling of transitions
of both the donor and acceptor molecules with the cavity mode leads
to the formation of three polaritonic branches, upper (UP), middle
(MP) and lower (LP). The vibrational bath causes transitions
between these modes, as well as to dark excitonic states,
allowing relaxation to the lowest energy states.

Treating the coupling to vibrational modes via Redfield theory
provides clear insights into mechanisms underpinning energy
transfer, but has limitations. Most notably, Redfield theory is
perturbative, holding in the limit of weak coupling to the
environment. In the occurrence of simultaneous strong coupling to
light and to vibrational modes, other methods are needed.
Indeed, recent work~\cite{rouse2024influencestrongmolecularvibrations}
has shown how results of Redfield treatments can break down in the
limit of strong coupling to vibrational modes. In addition to such
approaches being restricted to weak coupling to vibrational modes,
there are technical challenges with deriving the Redfield equation
for systems with strong light--matter coupling. The derivation of
the Redfield master equation requires decomposing the system--environment
interaction into terms which each evolve monochromatically---i.e.
eigenoperators of the system Hamiltonian. For interacting many-body
systems (such as saturable emitters coupled to a photon mode) this is
in general challenging, as it requires determining the full
eigenspectrum of the system.  As such, use of Redfield theory
generally requires additional restrictions. For example, one can
consider the few-excitation limit, where the model is equivalent
to coupled harmonic oscillators~\cite{Pino_njp17,MartinezMartinez_njp20},
or restrict to small numbers of emitters~\cite{saez_prb97}.

An alternative to the Redfield approach is to consider simplified
models, such as the Holstein--Tavis--Cummings (HTC) model, which
describes a single vibrational mode for each molecule~\cite{Cwik_2014}.
In such a model, the reduced size of Hilbert space makes it possible
to exactly treat the vibrational state without perturbative
approximations. This can be combined with mean-field~\cite{Strashko2018}
or cumulant~\cite{Arnardottir2020} approaches to describe the coupling
to cavity modes. However, while the HTC model can provide a reasonable
picture of sharply defined vibrational modes and their effects on e.g.
the absorption spectrum, it does not capture the effects of
energy relaxation associated with the continuum of low frequency modes.
Given the large number of such modes, it is not possible to exactly
include these in the system Hamiltonian~\cite{reitz_prres2}.

Here, inspired by a recent experiment~\cite{georgiou_ultralong-range_2021},
we explore the dynamics of a system of two molecular species coupled to
a common cavity mode, varying the corresponding coupling strength to
vibrational modes. To properly capture the vibrational bath, we rely on
the process tensor matrix product operator (PT-MPO) method combined with
a mean-field approach for the cavity
modes~\cite{fowler-wright_efficient_2022,fux_oqupy_2024} that allows
simulation cost independent of the number of emitters. We see that,
for weak vibrational coupling, our approach is in good agreement
with Redfield theory. That said, as the coupling to the vibrational
bath increases, the non-Markovian character of the dynamics becomes
more evident. Finally, when the coupling to the vibrational bath
becomes comparable to the light--matter coupling, we see a striking
change to the dynamics, which may potentially be attributed to the
formation of polarons. We thus demonstrate the importance of
introducing non-Markovian descriptions for the accurate design and
modeling of polaritonic cavities for energy-transfer applications. 

\section{Method}\label{sec:method}

\subsection{Model}
\begin{figure}
    \centering
    \includegraphics[width=0.5\textwidth]{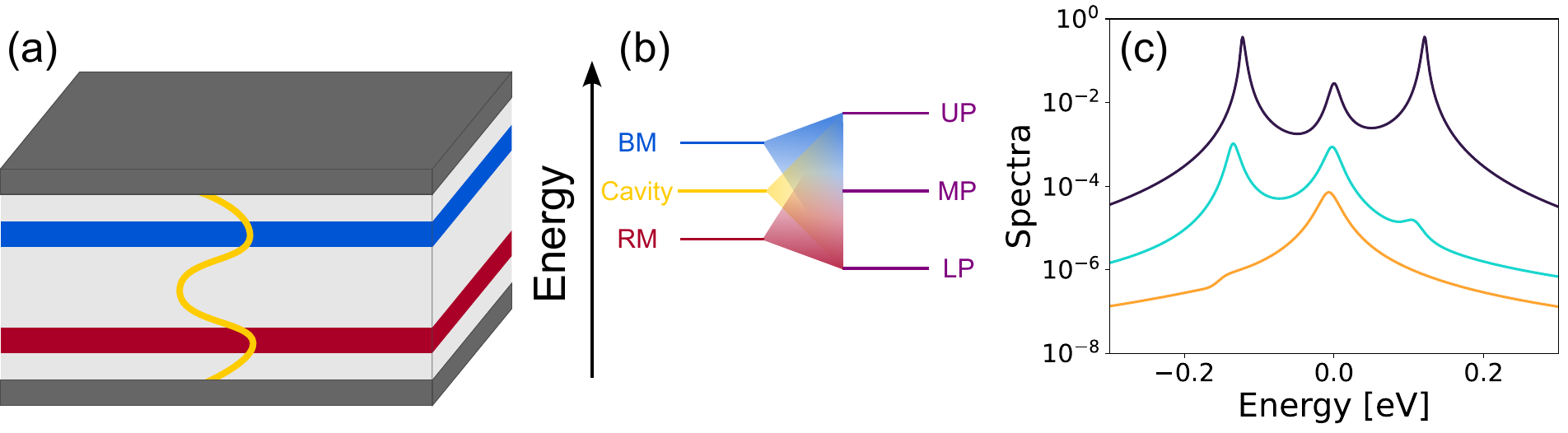}
    \caption{(a) Sketch of the physical system under consideration: a planar microcavity containing spatially separated layers of two different species of molecules, with lower (red) and higher (blue) transition energies. We consider a single photonic mode confined in the cavity.
    (b) A simplified energy level diagram of the system showing the relative positions of the exciton energy in the higher-energy blue molecule (BM) and the lower-energy red molecule (RM). We take the single cavity mode (cavity) to be at the average energy of the two species.  This leads to the middle polariton (MP) aligning with the original photon mode energy, and the upper and lower polaritons (UP and LP) lying above and below the original molecular energy levels respectively.
    (c) Time-dependent coherent spectra, $S(\omega,t)$, plotted at equal times ($t=500$ eV$^{-1}$)
    for three values of coupling strength to the vibrational bath, $\alpha$, corresponding to the three regimes of behavior discussed below; $\alpha$  increases from top (dark blue line) to bottom (orange line).
    The behaviors seen here are discussed at the start of Sec.~\ref{sec:results}.
    Other parameters used in (c) are $\kappa=0.01$ eV, $\Gamma_\downarrow=0.002$ eV,  $\frac{\Omega_1}{\sqrt{N_1}}=\frac{\Omega_2}{\sqrt{N_2}}=0.1, \omega_1-\omega_c=-(\omega_2-\omega_c)=0.1$ eV and the vibrational environment has parameter $\nu_c=0.16$ eV and temperature $k_B T=0.026$ eV.
    }
    \label{fig:sketch}
\end{figure}

Throughout the paper we consider a system consisting of two spatially separated layers of different species of molecules embedded into a microcavity, as shown in Fig. \ref{fig:sketch}(a). The two molecule species have their main transitions at different energies; we thus call the higher-energy molecule blue, and the lower-energy one red. 
We neglect energy disorder in the molecules, as well as direct dipole--dipole interactions between the molecules.
We also consider a single cavity photon mode coupled to both molecules, whose energy lies between those of the two molecules. These approximations and other possibilities are further discussed at the end of the paper.

The light--matter coupling strength is large enough to enable hybridisation of the three constituents into three polariton states: UP, MP, and LP, along with dark states corresponding to both molecular species. The molecules are not purely described by their electronic state, but also have a continuum of low-energy vibrational modes.  We treat these low-energy vibrational modes as a non-Markovian bath, coupling to the \emph{system} of photons and electronic states of the molecules.

Based on the discussion above, the system (i.e. excluding for now the bath) is modeled with a Tavis--Cummings Hamiltonian with $N=N_1+N_2$ emitters, where $N_s$ denotes the number of molecules of species $s$ (throughout the paper we use $\hbar=1$):
\begin{align}\hat H_S = \omega_c \hat a^\dagger \hat a + \sum_{s=1}^{2}\sum_{\text{\(n=1\)}}^{N_s}\left[\frac{\omega_s}{2}\hat \sigma_{s,n}^z + \frac{\Omega_s}{2\sqrt{N}}(\hat a\hat \sigma_{s,n}^+ + \hat a^\dagger\hat \sigma_{s,n}^-)\right].\label{eq:HS}
\end{align}
Here, $\omega_c$ is the frequency of the cavity photon mode and $\hat a$ ($\hat a^{\dagger}$) the corresponding annihilation (creation) operator, $\omega_s$ is the transition frequency of molecule species $s$, and $\hat \sigma_{s,n}^+$ $(\hat \sigma_{s,n}^-)$ is the raising (lowering) operator for the $n$th molecule of species $s$. The collective coupling of each species to light is $\Omega_s\sqrt{{N_s}/{N}}$.  Because our model (including the coupling to the bath discussed below) has a weak symmetry~\cite{albert2014} under the unitary rotation $U=\exp[i \theta (\hat a^\dagger \hat a + \sum_{s,n} \hat \sigma^z_{s,n}/2)]$, we can transform to a frame rotating at the cavity frequency by replacing $\omega_{1,2} \to \omega_{1,2}-\omega_c$ and setting $\omega_c=0$;  all results in this article are shown in this rotating frame.

The local vibrational bath for each molecule, and its coupling to the system, is described by:
\begin{equation}
    \hat H_E^{(s,n)} = \sum_k \left[\nu_k^{(s)} \hat b^\dagger_k \hat b_k + \frac{\lambda_k^{(s)}}{2}(\hat b_k + \hat b_k^\dagger )\hat \sigma_{s,n}^z \right],
\end{equation}
where $\hat b_k$ is the annihilation operator of a vibrational mode with energy $\nu_k^{(s)}$ for species $s$, and $\lambda_k^{(s)}$ the 
coupling strength to that mode. The couplings are fully 
characterized by the spectral density function $J_s(\nu)=\sum_k(\lambda_k^{(s)}/2)^2\delta(\nu-\nu_k^{(s)})$.
We consider Ohmic spectral densities for each species,
\begin{equation}
    J_s(\nu)= 2 \alpha \nu e^{-(\nu/\nu_c)^2},\quad \nu >0,
\end{equation}
which is chosen to capture the low frequency continuum of modes 
conferred by the host matrix of molecular aggregates~\cite{georgiou_ultralong-range_2021}.
Here $\alpha$ is a dimensionless coupling strength, and $\nu_c$ denotes a cutoff for this low-frequency behavior.    
As discussed in Ref.~[\onlinecite{fowler-wright_efficient_2022}], the parameters in such a spectral density can be fitted to match the absorption spectrum for some organic molecules---e.g. BODIPY-Br in Ref.~[\onlinecite{fowler-wright_efficient_2022}], for which $\alpha=0.25$ was found to be the relevant value.   
In the current work we are interested in exploring how the strength of coupling to vibrational modes affects the energy transfer process, and so we consider the Ohmic spectral density as a simple model expression and treat $\alpha$ as an adjustable parameter.
In the rest of this article we assume the same spectral density for both species, however this assumption can easily be relaxed.

In addition to the interaction with the vibrational baths, we include two Markovian dissipation channels: photon loss at rate $\kappa$, and non-radiative decay of
the molecule occupation
at rate $\Gamma_\downarrow$.
That is, we consider two additional contributions to the system density matrix evolution:
\begin{equation}
\partial_t \rho = -i[\hat H_S,\rho] + 2\kappa\mathcal{D}[\hat a,\rho] + \sum_{s,n}\Gamma_\downarrow\mathcal{D}[\hat \sigma_{s,n}^-,\rho]    
\end{equation}
where $\mathcal{D}[\hat X, \rho]=\hat X \rho \hat X^\dagger - \{ \hat X^\dagger \hat X, \rho\}/2$.
Note that the coupling to the vibrational bath causes both energy transfer processes---discussed further below---as well as pure dephasing. We assume that the dephasing of the molecules is fully accounted for by this coupling to the vibrational bath, and do not add any additional dephasing rate.
In the following we consider the dynamics of the system after both molecular species are assumed to be populated by a pump at time $t=0$, i.e.\ we time-evolve the system dynamics from an initial state of symmetric occupation of both red and blue molecules,
\(\rho^{ee}_{1,n}(0)=\rho^{ee}_{2,n}(0)=0.1\).
This initial state is similar to that prepared in the experiment that inspired this paper~\cite{georgiou_ultralong-range_2021}.
Because mean-field approaches cannot spontaneously break symmetry, we also add a small ``seed'' coherent field, \(\langle a(0) \rangle = 0.1\). An open question for future work is thus to consider approaches beyond mean-field, and explore how energy transfer processes are initiated in the absence of a symmetry-breaking field.
We also assume a thermal state for the vibrational environment.

\subsection{Mean-field PT-MPO}
To accurately calculate the dynamics in our model, we use the process tensor time evolving matrix product operator (PT-TEMPO) method.
As discussed in Refs.~[\onlinecite{pollock_non-markovian_2018,fux_efficient_2021,fux_oqupy_2024}],
the process tensor captures all possible effects of an environment on the system. 
This method provides a numerically exact description of the vibronic coupling.
While the PT-TEMPO method makes the calculation efficient for a single molecule, its direct application to multiple molecules would lead to a system Hilbert space growing exponentially with the number of molecules, rapidly becoming computationally intractable beyond a few molecules.
For this reason, and also to account for the coupling
to the common cavity mode, we use a mean-field approximation 
\cite{fowler-wright_efficient_2022}.
The mean-field approximation means assuming a product state
$ \rho = \bigotimes_{s,n} \rho^{(s,n)} \otimes \rho^{(c)} $
for the many-body density operator where $\rho^{(c)}$ is the density operator for the cavity mode and the other part of the state represents the molecules (see Refs.~[\onlinecite{fowler-wright_efficient_2022,fux_oqupy_2024}]
for details).  
For the cavity mode, the only information required is the expectation value \(\langle \hat a \rangle\).
This mean-field ansatz is known to become exact for the many-to-one coupling described by Eq.~\eqref{eq:HS} as the number of emitters \(N\to\infty\).~\cite{mori_exactness_2013,carollo_exactness_2021}
For the current system, this reduces the problem to three coupled systems: the cavity mean-field \(\langle \hat a \rangle\) interacting with two representative molecules,
$\rho_{1}$, $\rho_{2}$,
since all molecules of a given species are identical and therefore have the same on-site properties.
In this way we are able to describe the two-species--cavity
system efficiently for large (e.g.\ $N\sim 10^8$) numbers of molecules.

The mean-field approach with multiple molecular species was recently implemented with the PT-TEMPO method in the OQuPy Python package \cite{gerald_e_fux_tempocollaborationoqupy_2024}.
The PT-TEMPO method implemented here can readily include the Markovian Lindblad terms mentioned above in addition to the non-Markovian treatment of the vibrational environment.
We note that the mean-field approach used here does not have any restriction to small numbers of excitations: in principle the calculations we present can be used at arbtirary excitation density.
If one is operating at weak excitation densities, there are simpler approaches one may consider.
These include working in a restricted Hilbert space, i.e the single excitation subspace~\cite{scholes_rsa476}.
In addition, as recently discussed~\cite{schwennicke2024}, for low excitation density one can use linear optical response---and in particular optical filtering---to understand how the polariton system responds to an external pump, including in the time domain.

\subsection{Time-dependent coherent spectra}
To extract information about the dynamics of the system, we look at the \emph{time-dependent coherent spectra},
\begin{equation}
    S(\omega,t) = \left|\int_{-\infty}^{\infty} w(t-t',\Delta t)\langle \hat a(t')\rangle e^{-i\omega t'} dt'\right|^2,
\label{Eq:spectra}
\end{equation}
where $w(t,\Delta t)$ is an appropriate time window function---effectively a top-hat function extending over $t<t'<t+\Delta t$.
In Appendix~\ref{App:Spectra} we discuss how we implement this window function following an approach introduced in Ref. [\cite{mark_spectral_1970}].   Throughout this article we use $\Delta t=20$ eV.

In general, the spectrum involves a two-time correlation function; however, within the mean-field approach introduced above, the two-time correlation function of photons factorizes. As such, within this approximation we directly perform the windowed Fourier transform on the coherent field expectation. 
This leaves us with information on coherent processes and their contribution to the spectra.
For the initial conditions we consider, and considering the limit of large $N$, this approach can be reasonable.  
There may however be finite $N$ effects that could modify the shape (e.g. linewidth) of the spectra $S(\omega, t)$ beyond the mean-field description presented here---particularly if one were to extend the model to include multiple cavity modes.
This mean-field approach does, however, provide important information regarding the eigenenergies of the system, as well as the occupation of the different eigenstates. 

\section{Results and Discussion}\label{sec:results}

\subsection{Dependence of spectrum on coupling to vibrational bath}

Figure~\ref{fig:sketch}(c) shows examples of the coherent time-dependent power spectrum, $S(\omega,t)$, for a fixed time at three different values of the  coupling strength to the vibrational bath,  $\alpha$.
These three values are chosen to illustrate the three different coupling regimes that we find arise, and which we discuss in detail below.
For the lowest alpha, (dark blue line, top) three clear resonances are seen, corresponding to UP, MP, and LP.
As we discuss below, this value of $\alpha$ corresponds to the Markovian regime, where approximate methods such as Redfield theory match the exact results well.
When we increase $\alpha$ (light blue line, middle) we see that the UP peak has almost disappeared; it is several orders of magnitude smaller than the LP and MP resonances.   This reflects the increased rate of vibrationally-induced energy transfer.
%, where we see a transfer to the lower polariton state from the higher energy states with time.
As discussed further below, this value of $\alpha$ is such that Markovian approximations fail, and so we label this the non-Markovian regime.
Finally, the orange line shows a spectrum corresponding to an even higher $\alpha$;  here yet different behavior is seen, with a suppression of relaxation into the LP mode, leading to a spectrum with a single peak near zero energy. As will be discussed below, we suggest that one may associate this regime with the formation of polarons.
It is also notable the overall amplitude of the spectrum decreases with increasing $\alpha$;
as we discuss below, this is due to vibrationally-induced
transfer to dark excitonic states.
% since the coupling to the vibrational bath conserves %excitation number, there is no direct effect of $\alpha$ on loss rate.  
%However, the vibrationally-induced transfer of excitations between different modes can change the overall loss rate.  
%The cavity decay rate $\kappa$ is assumed faster than the loss rate of the molecules $\Gamma_\downarrow$.  For the parameters considered here, the MP has a larger photon fraction than the LP and UP, and as such, transfer to the MP leads to a faster overall decay rate.
We note here that, by comparing our calculated spectra with experimental ones~\cite{georgiou_ultralong-range_2021}, 
we estimate that representative values of $\alpha$ for several commonly used molecules are of the order of $0.15 - 0.20$, implying that the entire range
of coupling strengths covered in this work is experimentally feasible.

\begin{figure}
    \centering
    \includegraphics[width=0.5\textwidth]{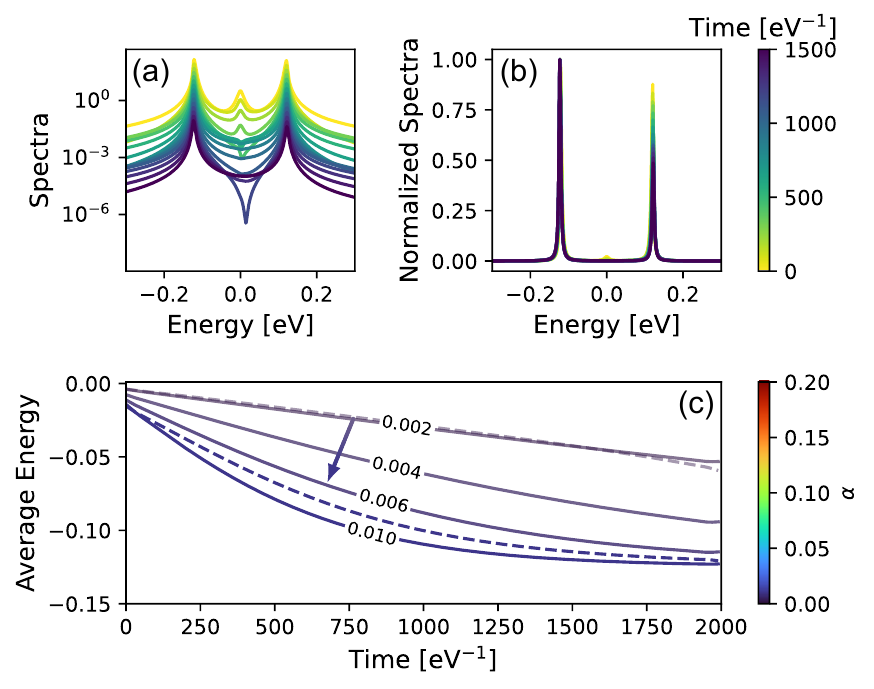}
    \caption{
    Dynamics in the Markovian regime.
    Panels (a) and (b) both show the time-dependent coherent  spectra, $S(\omega,t)$ plotted vs frequency at a set of times indicated by color (see color scale on the right) for $\alpha=0.002$. 
    In (a) the raw spectra are plotted on a semilog scale, while in (b) the same data are plotted on a linear scale, normalized so the maximum value is $1$.
    (c) Average energy as a function of time (see text for details) for a range of (small) values of $\alpha\in[0.002,0.01]$, with values as indicated by the color scale on the right.  $\alpha$ increases from top to bottom as indicated by the arrow.
    Dashed lines show the corresponding predictions of the weak-vibrational-coupling Redfield theory described in Appendix~\ref{app:Redfield}, for the highest and lowest $\alpha$ in the range. Other than $\alpha$, all parameters take the same values as in Fig.~\ref{fig:sketch}(c).
    }
    \label{fig:Spectra1}
\end{figure}

In Fig.~\ref{fig:Spectra1} we show in more detail what happens at small values of coupling $\alpha$. Panels (a) and (b) show how the spectrum evolves with time.
Initially there is high occupation in all three polariton modes. 
As time evolves two effects occur:  all three peaks decay with their own individual decay rate (dependent on the exciton and photon fraction in each mode), and in addition, there is vibrationally-driven population transfer.
Regarding the decay of peaks, it is important to note that the overall spectrum is clearly decaying with time; this is visible on panel (a) where we show the un-normalized (raw) spectrum on a semi-log scale, so that decay is visible by the downward shift of successive lines.  Panel (b) shows a normalized spectrum so that overall decay is not visible, only relative differences of decay rates between different modes.

The vibrationally driven population transfer arises
because the operator coupling each molecule
to the vibrational bath, \(\hat{\sigma}^z_{s,n}\), 
does not commute the system
Hamiltonian \(\hat{H}_s\). Consequently, the system-bath interaction
causes scattering between eigenstates.
Note \(\hat{\sigma}^z_{s,n}\) does commute with the total number of excited molecules plus photons, so the vibronic coupling does not 
lead to overall loss of excitations. However, one must consider that in addition
to the bright states (upper, lower and middle polaritons), there are
a large number of optically dark states at the bare excitonic energies.
Scattering into these dark states due to vibronic coupling results in a reduced spectral output from the bright modes. In Appendix~\ref{sec:dark_bright_exciton_populations} we 
provide expressions for the bright and dark state populations
in our model and evidence that
increasing \(\alpha\) leads to increasing depletion of the bright populations into the dark modes. It is worth noting that there is only
effective transfer from the dark modes back to the bright states at early times, driven by the transient coherence of the bright states induced by the initial state.  
The absence of dark to bright transfer in the absence of coherence is an established result~\cite{schwennicke2024} in the mean-field limit.  Note however that the small effective transfer rate ($1/N$ in the number of molecules) can be amplified when there is a large dark state population.
%
% Regarding the vibrationally-driven population transfer, this can be understood as a consequence of the system--bath coupling not commuting with the bare system Hamiltonian.  
% This means that the system--bath coupling can cause scattering between different kinds of excitation.  
% It does however commute with the total number of excited molecules plus photons, so this coupling to vibrational modes cannot cause loss of excitations, just interconversion between the different states (upper, lower, middle polaritons and dark excitons).
% Since we only plot observables that involve the cavity photon field, scattering into dark exciton states would however appear as an effective loss mechanism within the figures we plot.

As discussed below, for small $\alpha$, the behavior observed matches that described by the Redfield theory, and one can understand such process from the Redfield equation given in Appendix B. What one sees from that is indeed that the dissipation terms can scatter excitations between different states.  Such behavior---in the limit of weak coupling to vibrational modes---has been discussed elsewhere in some detail[\cite{Pino_njp17,MartinezMartinez_njp20,saez_prb97,saez_aom8}].

As noted above, the bare decay rate of the MP is faster than that of the LP and UP.
For the parameters used in Fig.~\ref{fig:Spectra1},
vibrationally-driven transfer to the dark
states is slower than this bare decay rate
and so the MP decays the fastest.
%the vibrationally-driven transfer is not fast enough to compensate for the different decay rates and so the MP decays fastest.  In the linear-scale plot (panel b) one however sees that at late times the UP population has decreased compared to the LP---this reflects the vibrationally-driven transfer process.
In the semilog-scale plot (panel a), one may see that at late times the MP peak is in fact replaced by a dip in the spectrum.
This feature corresponds to a Fano-like~\cite{fano1961} effect, arising due to interference between the different contributions to the spectrum: the definition of $S(\omega,t)$ involves a modulus squared of the contributions of the different peaks, which can arise with different phase factors.  As such---especially when the amplitudes of the different peaks are very different---this can lead to dips and asymmetric features in the spectrum.
Given long enough time, the LP will fully dominate the spectra, but for such a low $\alpha$ that only happens at times where the total value for the spectra is very low.

To further characterize the energy transfer process, it is instructive to look at the average energy---i.e.\  $\int d\omega \, \omega S(\omega,t)/\int d\omega \, S(\omega,t)$---as a function of time. 
Panel (c) shows how this quantity behaves with time for a range of values of $\alpha$, all in the small $\alpha$ regime. 
As our initial condition is an equal occupation of the molecules and the energies $\omega_{1,2}-\omega_c$ are symmetric around zero, the average energy should start at zero. However, because our time window is defined to start at $t$ and extend forward to $t+\Delta t$, the average energy at $t=0$ can already deviate from being exactly at zero.
In the limit $\alpha\to0$ there would be no route to vibrational population transfer. Then, because the LP and UP modes have equal photon weight these peaks would decay equally fast, causing the average energy to remain at zero.
A small vibrational coupling of $\alpha=0.001$ already gives a channel for energy relaxation, which produces a slow decrease of the average energy, shown as the darkest blue line. 
Increasing $\alpha$ increases the rate of relaxation. 
For large $\alpha$ we see the energy relaxation saturates, corresponding to a final state that is almost entirely the LP.
For the values of $\alpha$ shown in Fig.~\ref{fig:Spectra1} the same behavior can also be recovered from the Redfield theory~\cite{Pino_njp17,MartinezMartinez_njp20} described in Appendix~\ref{app:Redfield}---this is shown by dashed lines in the figure.

\begin{figure}
    \centering
    \includegraphics[width=0.5\textwidth]{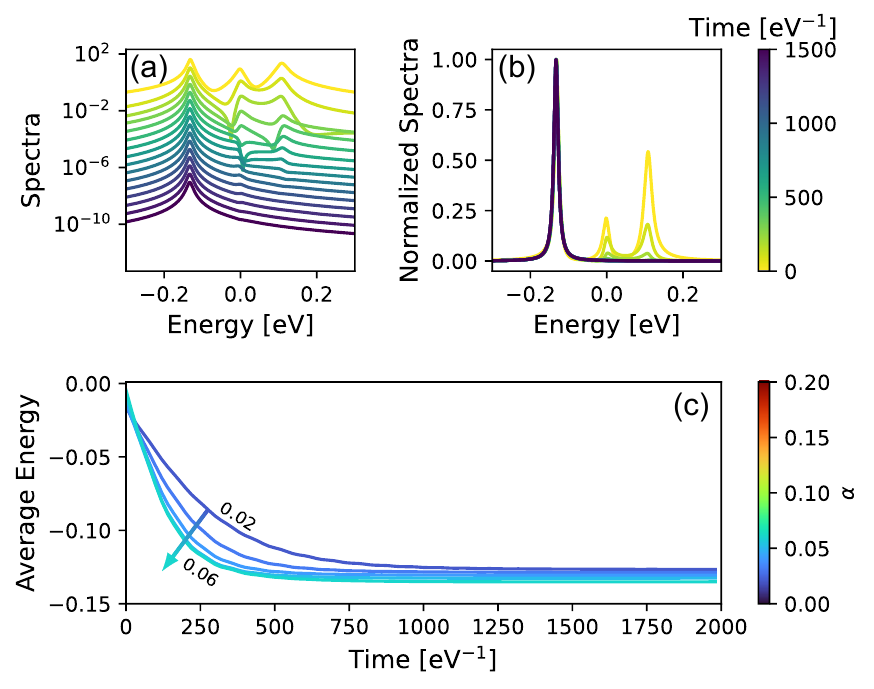}
    \caption{%\ct{we say panels are same as Fig. 2... are we going to show Redfield dashed lines failing?}
    Dynamics in the intermediate $\alpha$ (non-Markovian) regime.  The panels are the same as for Fig.~\ref{fig:Spectra1} but for larger values or ranges of $\alpha$.  
    Panels (a) and (b)  plotted for $\alpha=0.05$, and panel (c) for $\alpha\in[0.02,0.06]$. 
    Other than $\alpha$, all parameters take the same values as in previous figures.}
    \label{fig:Spectra2}
\end{figure}

Figure~\ref{fig:Spectra2} shows results for a range of larger values of vibrational-coupling strength $\alpha$.
For these values, the results no longer match well to the Redfield theory---the actual rate of energy relaxation is larger than the Redfield theory would predict.
% in a different regime, which is beyond Markovian and exhibits efficient polariton mediated transfer. 
% When we increase $\alpha$ we see that the energy drops down to the lower polariton energy faster, and faster than the Markovian description predicts. 
In this case different behavior is seen in the evolution of the spectrum, i.e. in panels (a) and (b).
This shows a fast depletion of the UP state, and subsequent depletion of the MP, leading to a single peak at the LP energy for late times. 
Panel (a) again shows Fano-like features associated with the depleted LP and MP peaks.
The average energy in panel (c) shows similar behavior to that seen at the larger values of $\alpha$ in Fig.~\ref{fig:Spectra1}---there is rapid relaxation to to an average energy consistent with only populating the LP state. 
This non-Markovian regime is exactly where the energy transfer become the most efficient; experiments would thus benefit from pursuing vibrational coupling strengths in this regime.

\begin{figure}
    \centering
    \includegraphics[width=0.5\textwidth]{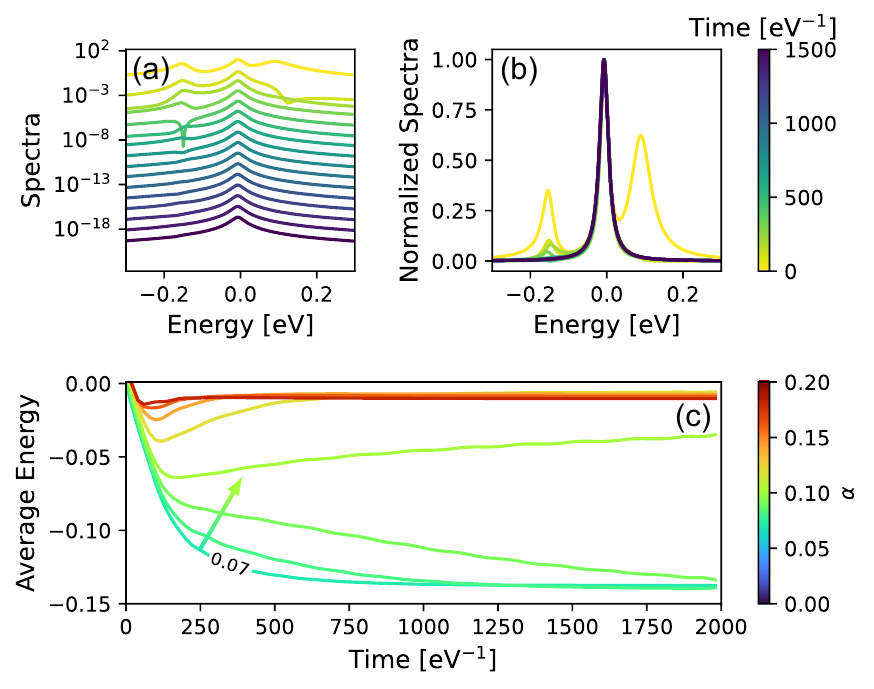}
    \caption{Dynamics in the large $\alpha$ regime.  The panels are the same as for Figs.~\ref{fig:Spectra1},\ref{fig:Spectra2} but for even larger values or ranges of $\alpha$.  
    Panels (a) and (b)  plotted for $\alpha=0.14$, and panel (c) for $\alpha\in[0.07,0.18]$.
    Other than $\alpha$, all parameters take the same values as in previous figures.
    }
    \label{fig:Spectra3}
\end{figure}

When the vibrational-coupling strength $\alpha$ is even larger, the energy transfer process breaks down, as can be seen in Fig. \ref{fig:Spectra3}.
For $\alpha \simeq 0.1$ one still sees (in panel (c)) the average energy decrease toward that of the lower polariton.
However for the largest $\alpha$ the energy transfer process appears to cease, with the late-time behavior showing an average energy near zero. 
An example of this is shown in panels (a) and (b) for $\alpha=0.14$, where one sees a spectrum which shows a single peak near the bare MP (or bare photon) peak.  

At sufficiently large coupling to vibrational modes, it is known~\cite{Herrera2016cavity,zeb2018exact,Wu2016pp,rouse2024influencestrongmolecularvibrations} that one can have an effective ``transition'' where polaron formation causes suppression of polariton formation.
In such a transition, the formation of polarons (i.e. dressing of the molecular exciton by the displaced vibrational modes) leads to a strong suppression of the effective light--matter coupling due to the reduced overlap of the vibrational state in the ground and excited electronic states. 
Were such an effect to arise, one would see an emission spectrum corresponding approximately to the bare cavity mode, corresponding to cavity-filtered emission by the molecules.  The behavior seen  in Fig.~\ref{fig:Spectra3} is potentially consistent with this scenario.

However, a number of features suggest a more complex explanation may be required for these results.
First, one may note that apparent LP and UP peaks are observed at early times.
This could  be associated with the finite time required for polaron formation, a question which could be probed in future work by exploring the dynamics of the vibrational reservoir using methods described in Refs.~[\onlinecite{fux_oqupy_2024,Gribben2022usingenvironmentto}].
Second, the behavior changes when cavity loss is further increased.
These additional results (discussed below) may suggest that the peak seen in Fig.~\ref{fig:Spectra3} is still the MP peak, in this case the behavior observed should be associated with a regime where polariton formation persists, but potentially transfer between modes becomes suppressed as one approaches the polaronic regime.

\subsection{Dependence of spectrum on cavity loss}

\begin{figure}
    \centering
    \includegraphics[width=0.5\textwidth]{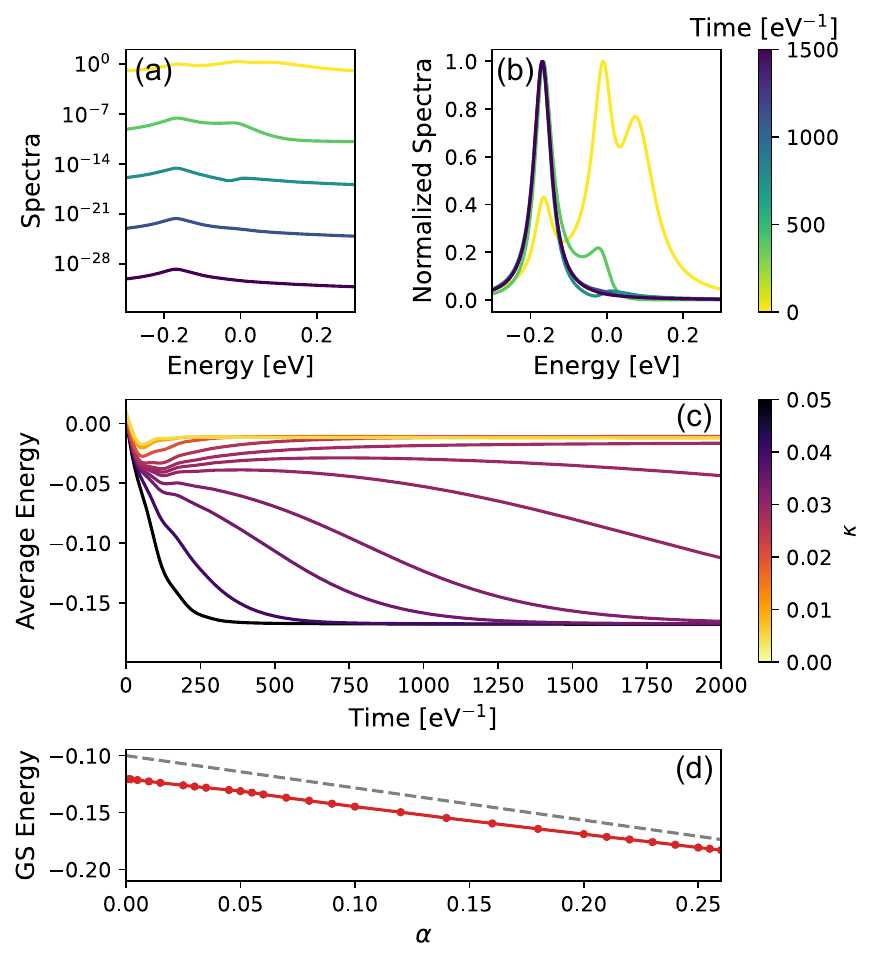}
    \caption{Dynamics for larger cavity loss rate, $\kappa$.
    (a, b) Normalized spectra for a high vibrational bath coupling, $\alpha=0.25$, on semi-log (a) and linear (b) scale. Different colors correspond to different times, as indicate by the color bar on the right.
    (c) Average energy as a function of time, for constant $\alpha = 0.25$ and varying $\kappa$ as indicated by the color bar on the right-hand side.
    (d) Location of the lowest energy peak (i.e. maximum) in the spectra at late times ($t=1500/eV$) for different $\alpha$ (solid line). The dashed line is the expected exciton energy of the red molecule shifted by the reorganization energy.
    }
    \label{fig:LargeKappa}
\end{figure}

Figure~\ref{fig:LargeKappa} shows how the behavior changes when we increase the cavity loss rate.
As noted earlier, the behavior observed in the spectrum has two (potentially competing) processes occurring:   energy transfer driven by the vibrational coupling, and the different decay rates (due to different composition) of the different collective modes.
Increasing $\kappa$ further increases this latter effect---i.e. causing yet faster decay of the MP relative to the LP and UP peaks.
Figure~\ref{fig:LargeKappa}(a) shows that the combination of large $\alpha$ and large $\kappa$ then leads to a two-stage dynamics. We first see an evolution toward a single peak near zero energy (as seen in Fig.~\ref{fig:Spectra3}).  
However, at later times this then evolves into a lower energy peak.  

If polariton modes are still well defined, this can be explained by the increasing $\kappa$ causing faster decay of the MP, thus revealing the LP peak at late times.
Alternatively, this late time behavior may be the result of cavity-filtered emission from the low energy molecules, analogous to effects that have been seen in other contexts~\cite{Hughes2011influence}.

We note that one can indeed see signatures of strong polaronic shifts of the excitons in this large $\alpha$ regime.   
This is seen in Fig.~\ref{fig:LargeKappa}(b), which shows the location of the lowest energy peak seen in the spectrum at late times (for $\kappa=0.05$ eV), demonstrating that the peak energy decreases with increasing $\alpha$.
For comparison, the energy of the vibrationally relaxed red-molecule exciton energy is also shown.
The emission frequency of the low energy peak closely tracks the vibrationally relaxed energy of the red molecules, although is consistently below this energy.

\section{Discussion and conclusion}
In conclusion, we have calculated the dynamics of a system of two species of organic molecules strongly coupled to a common cavity mode.
We showed how the strength of the coupling of the molecules to their respective vibrational baths qualitatively changed the energy relaxation dynamics, and thus the time evolution of the emission spectra. For a high rate of energy transfer, the dynamics differs significantly from that predicted by a Markovian Redfield theory.  
Furthermore, when vibrational coupling  becomes too strong,  energy transport can potentially be disrupted by the formation of polarons which would ultimately suppress the light-matter coupling.
The values of $\alpha$ at which this change of behaviour occurs are comparable to those found to be relevant for some organic materials in Ref.~[\onlinecite{fowler-wright_efficient_2022}].

To perform these calculations we applied a recently-developed~\cite{fux_oqupy_2024} method for calculating dynamics of a quantum system with multiple species of emitters in the mean-field limit, using a PT-MPO method for capturing the effects of a the non-Markovian bath. Our work clearly shows the potential of using such methods to probe dynamical effects such as energy transfer in  systems with strong light--matter coupling.

There are a number of restrictions or simplifications that were required in our work to reach tractable results.
Most notably, we consider a single cavity mode, in contrast to the multiple different in-plane modes present for a planar cavity~\cite{Arnardottir2020,Menghrajani2024}.
As discussed in Ref.~[\onlinecite{Arnardottir2020}], there is a close link between a single-mode approach and the use of mean-field theory: considering more cavity modes provides a larger contribution from beyond-mean-field fluctuations.  Vice versa, mean-field theory is good if only a single momentum state is macroscopically occupied, in which case it can be reasonable to restrict the model to just that state.  
While beyond-mean-field approaches have been used with simpler models of vibrational modes~\cite{Arnardottir2020,Menghrajani2024}, combining these with the continuum of vibrational modes considered here is an open challenge for future work.

There are further features of experiments not included in the current work, although the methods here could in principle be extended to address these.
Energetic disorder of the molecular species has not been included---the only sources of broadening in our model are homogeneous terms such as decay and coupling to the vibrational modes. 
In principle disorder could be included by replacing each molecular species by a set of many separate species, with each one differing by small changes in the molecular energy.  
This would however considerably increase the computational cost of the calculation.
We also considered only one form of spectral density, describing a broad low-frequency distribution; some materials may however show spectral densities including sharp vibrational modes.
The TEMPO approach to deriving process tensors is best-suited to broad spectra~\cite{fux_oqupy_2024}, however other methods for deriving process tensors~\cite{cygorek2022simulation} can deal with sharp features.

By accessing parameter regimes beyond the weak vibrational coupling approximation, our results show that there exists a ``sweet spot'' in vibrational coupling that maximizes energy transfer rates.  One clear question for future work is to probe how the state of the vibrational environment evolves, and whether there are signatures in the dynamics of polaron formation suppressing polariton formation.
Another broad question is to explore how one may identify molecular systems that reach the optimal parameters for energy transfer, and how this understanding can be exploited to design systems to controllably funnel energy into desired modes.
The model we consider has a larger parameter space in terms of mode energies, detunings, properties of vibrational spectra, and future work may systematically explore the optimal conditions in this large parameter space.

\section{Acknowledgements}
We want to thank David G. Lidzey, Stéphane Kéna-Cohen and Joel Yuen-Zhou for useful discussions.
K.B.A, B.W.L., and J.K. acknowledge support from EPSRC (Grant No. EP/T014032/1).
P.F.-W. acknowledges support from EPSRC (Grant No. EP/T518062/1)
The Center for Polariton-driven Light--Matter Interactions (POLIMA)
is sponsored by the Danish National Research Foundation
(Project No.~DNRF165).

\bibliography{JCPreferences}

\appendix

\section{Details of spectrum calculation}
\label{App:Spectra}
To calculate the time-dependent coherent spectra we use a top-hat window function of width $\Delta t$, which leads to Eq. (\ref{Eq:spectra}) taking the form
\begin{equation}
    S(\omega,t) = \left|\int_t^{t+\Delta t}\langle \hat a(t')\rangle e^{-i\omega t'}dt' \right|. \label{eq:S_App}
\end{equation}
When using a discrete Fourier transform, this time-window will naïvely lead to a limited frequency resolution, (${2\pi}/{\Delta t}$). One way to get around this is to numerically perform the integral over a larger window, adding 0 values outside of the window function. We use a variant of this approach\cite{mark_spectral_1970}, and evolve our photon field for a time beyond the time values we are interested in studying, and then perform the integral as
\begin{multline}
    \int_t^{t+\Delta t}\langle \hat a(t')\rangle e^{-i\omega t'}dt' 
    \\\approx \int_t^{T_\infty}\langle \hat a(t')\rangle e^{-i\omega t'}dt' - \int_{t+\Delta t}^{T_\infty+\Delta t}\langle \hat a(t')\rangle e^{-i\omega t'}dt',
\end{multline}
where $T_{\infty}$ is chosen such that the value of $\langle \hat a(t)\rangle$ is negligible for $t\in[T_{\infty},T_{\infty}+\Delta t]$. Using discrete Fourier transforms to approximate the two separate Fourier integrals on the right hand side of this expression then leads to a higher resolution of the resulting spectra.

As explained in the main text, Eq.~\eqref{eq:S_App} provides
information on coherent processes and their contribution to the spectra. See
Ref.~[\onlinecite{eberly_time-dependent_1977}] for further discussion of 
time-dependent spectra and the connection to the physical spectrum 
of light, i.e. that detected by a filtered measurement in experiment.\\

We should note that the form of the time-window function can lead to un-physical oscillations. To resolve this issue we introduce a low-pass filter in our calculations of the average energy in Figs. \ref{fig:Spectra1}-\ref{fig:LargeKappa} to filter out high frequency oscillations.

\section{Redfield Theory}
\label{app:Redfield}
%\subsection{Redfield Theory}
To gain an understanding of the role of non-Markovianity in the behavior we see, it is useful to compare our results with a Markovian alternative.
In studying the energy transfer it is important to fully capture rates of decay and transfer between modes. 
For that reason we want to avoid the secular approximation and so use Redfield theory as our Markovian comparison, (combined with a Lindblad term to capture cavity photon loss and non-radiative/non-vibrational loss in the molecules as discussed in the main text). 
As noted in the introduction, Redfield theory requires knowing the eigenoperators of the system Hamiltonian which is challenging for the Dicke model.
To overcome this we derive the transition rates by replacing spin operators by bosonic operators, which relies on the assumption that the molecular modes are only weakly occupied~\cite{Pino_njp17,MartinezMartinez_njp20}. 
This yields a model of non-interacting Bosons (the Hopfield model~\cite{Hopfield1958}) which can be fully diagonalised.
The numerical results we present are in the regime where this holds, and as the occupation fraction never exceeds the initial value, which we set to 10\% in our calculations.

Using these approximations, the resulting Redfield theory takes the form:
\begin{multline}
    \partial_t \rho = -i[\hat H_S,\rho] + 2\kappa\mathcal{D}[\hat a,\rho] + \sum_{s,n}\Gamma_\downarrow\mathcal{D}[\hat \sigma_{s,n}^-,\rho]
    \\+   \sum_{s,n}\left([\hat{\mathcal{A}}_{s,n}\rho,\hat{\mathcal{B}}_{s,n}] - [\rho\hat{\mathcal{A}}^\dagger_{s,n},\hat{\mathcal{B}}_{s,n}] \right)
\end{multline}
where $\kappa$ is the cavity loss rate, $\Gamma_\downarrow$ is the non-radiative decay (from other sources than the vibrational bath) and $H_S$ is the system Hamiltonian in Eq.~\eqref{eq:HS}. The effect of the vibrational mode is captured by the operators
\begin{eqnarray*}
    \hat{\mathcal{A}}_{s,n}&=&\sum_{\alpha\beta}C_{s,n,\alpha}^\ast C_{s,n,\beta} \gamma(\Delta_{\alpha\beta}) \hat{X}_{\alpha}^\dagger \hat{X}_{\beta} \\
    \hat{\mathcal{B}}_{s,n}&=&\sum_{\alpha\beta}C_{s,n,\alpha}^\ast C_{s,n,\beta} \hat{X}_{\alpha}^\dagger \hat{X}_{\beta}
\end{eqnarray*}
where the sums of $\alpha, \beta$ go over the three polariton branches (upper, lower and middle) as well as the $N-2$ dark states, and (as in the main text), $s$ labels the two species and $n$ indexes the sum over molecules.  One may note that $\hat{\mathcal{B}}_{s,n}$ corresponds to the bare system--bath coupling operator, $\simeq\hat\sigma^z_{s,n}$, and is thus Hermitian.
$C_{s,n,\alpha}$ is the Hopfield coefficient relating molecule $n$ of species $s$ to the mode corresponding to the operator $\hat{X}_{\alpha}$, and these operators are the eigenoperators, $[\hat{X}_{\alpha},\hat H_S]= \epsilon_\alpha \hat X_{\alpha}$,
of the system Hamiltonian (i.e.\ of the non-interacting Hopfield model). 
The energy difference of eigenoperators $\alpha$ and $\beta$ is denoted by $\Delta_{\alpha\beta}=\epsilon_\alpha-\epsilon_\beta$ and $\gamma(\Delta)$ is defined as: 
\begin{equation}
    \gamma(\Delta)=\int_{-\infty}^0e^{-i\Delta t}\phi(-t) dt
\end{equation} 
with $\phi(t)$ being the bath autocorrelation function\cite{Pino_njp17, MartinezMartinez_njp20}. This becomes
\begin{equation}
    \gamma(\Delta) =  
    \frac{i}{\pi} 
    \int d\omega  \frac{S(\omega)}{\Delta - \omega -i 0} 
    =
    S(\Delta)
    + \frac{i}{\pi}\int d\omega P \frac{S(\omega)}{\Delta - \omega} 
\end{equation}
where $S$ is the bath noise-power spectrum
\[
S(\omega) = \begin{cases}
    \pi J(\omega)[n_B(\omega)+1] & \omega\geq 0,\\
    \pi J(-\omega)n_B(-\omega) & \omega < 0.
\end{cases}
\]
with $n_B(\omega)$ denoting the Bose--Einstein occupation function.

To capture the dynamics of the photon field under the mean-field approximation we write down the equations of motion of the expectation values of the polariton operators:
\begin{multline}
    \label{Eq:Pol_eoms}
    \partial_t \langle\hat{X}_{\nu}\rangle 
    =
    -i\epsilon_{\nu} \langle\hat{X}_{\nu}\rangle 
    - \kappa  \sum_\alpha C_{c,\nu}^\ast C_{c,\alpha} \langle\hat{X}_{\alpha}\rangle +{}
    \\
    \sum_{i\alpha}C_{s,\nu}^\ast  C_{s,\alpha}\left(-\gamma(0) - \frac{\Gamma_\downarrow}{2} + (\gamma(\Delta_{\alpha s})^\ast -\gamma(\Delta_{s\alpha}))S_s \right) \langle\hat{X}_{\alpha}\rangle +{}
    \\
    \sum_{i\alpha\beta\delta}(\gamma(\Delta_{\beta\alpha})^\ast 
    -
    \gamma(\Delta_{\alpha\beta}))C_{s,\nu}^\ast C_{s,\alpha}^\ast  C_{s,\beta}C_{s,\delta} \langle\hat{X}_{\alpha} \rangle^\ast \langle\hat{X}_{\beta}\rangle\langle\hat{X}_{\delta}\rangle,
\end{multline}
where the sums over Greek indices now go over the three polariton modes.
When considering only these bright modes, the Hopfield coefficients $C_{s,n,\alpha}$ is independent of molecule index $n$ so we have suppressed this label and written $C_{s,\alpha}$ in the equation above.  We have also introduced $C_{c,\alpha}$, the Hopfield coefficient relating the cavity mode to the operator $\hat X_\alpha$,
and  $S_s = \frac{1}{N}\sum_n \langle\hat\sigma_{s,n}^+\hat\sigma_{s,n}^-\rangle$,  the normalized occupation of molecules of species $s$, which evolves according to
\begin{equation}
    \partial_t S_s = \sqrt{2}\Omega \sum_{\alpha\beta}\text{Im}\left[C_{c,\alpha}^\ast C_{s,\beta}\langle\hat{X}_{\alpha} \rangle^\ast \langle\hat{X}_{\beta}\rangle\right] -\Gamma_\downarrow S_s.
    \label{Eq:S_eom}
\end{equation}
The dynamics of the photon field can then be calculated by solving Eqs.~(\ref{Eq:Pol_eoms}-\ref{Eq:S_eom}) for a given initial state, and converting the polariton operator expectations back into the photon expectation. The spectra are then calculated in the same way as is done for the photon field found from the mean-field PT-TEMPO approach.

\section{Bright and dark exciton populations}
\label{sec:dark_bright_exciton_populations}
In this section we derive expressions for the bright and dark exciton populations
in the model, and show their dynamics during the energy transfer process through
the Markovian and non-Markovian regimes. 

For a model of a single photon mode interacting with \(N_1\) molecules of
one type and \(N_2\) molecules of a second type, there are two optically
`bright' exciton modes which couple with the cavity mode to form the lower,
middle and upper polaritons. 
This leaves a large number of `dark' exciton modes,  \(N_1-1\) of the first
type of molecules and \(N_2-1\) of the second type, which are orthogonal to the 
bright modes and at the molecular energies \(\omega_1\) and \(\omega_2\), respectively~\cite{zeb2022}.

As discussed elsewhere~\cite{herrera2017,herrera2017a,herrera2018,cwik2016}, 
in the presence of vibronic coupling, the dark modes become optically active.
This leads to transfer to the dark states and further a small spectral feature
at the exciton energies~\cite{cwik2016,fowler-wright_efficient_2022}. However,
the latter is not resolved in the results shown in the main text due to the large spectral weight of
the polaritons at these energies.  As such, for those figures (corresponding to what can be accessed experimentally), transfer to the dark states is evident only through its contribution to loss from the bright states.

We now derive expressions for the bright and
dark state populations within the mean-field description, and then present 
results for the dynamics of these populations during the energy transfer experiments.

Following Ref.~[\onlinecite{zeb2022}], we identify the bright  excitonic operator for
species \(s\) (\(s=1,2\))
\begin{align}
    \mathcal{B}_{s}^+ = \frac{1}{\sqrt{N_s}}\sum_{i=n}^{N_s} \sigma_{s,n}^+
\end{align}
The bright state population for \(s\) is then calculated in mean-field theory  as~\cite{fowler-wright_efficient_2022}
\begin{align}
    P_{s}^B &= \langle \mathcal{B}_{s}^+ \mathcal{B}_{s}^- \rangle \\
    &= \frac{1}{N_s} \sum_{n,m=1}^{N_s} \langle  \sigma_{s,n}^+  \sigma_{s,m}^- \rangle \\
    &=  \frac{1}{N_s} \left[ 
    \sum_{n=1}^{N_s} \langle  \sigma_{s,n}^+  \sigma_{s,n}^- \rangle
    + 
    \sum_{\substack{n,m=1\\n\neq m}}^{N_s} \langle  \sigma_{s,n}^+  \sigma_{s,m}^- \rangle
    \right] \\
        &\approx  \frac{1}{2}\left( 1 + \langle  \sigma_{s,n}^z  \rangle\right)
    + (N_s-1) \langle  \sigma_{s,n}^+ \rangle \langle \sigma_{s,m}^- \rangle.
    \label{eq:PBs1}
\end{align}
Here in the final line we used the mean-field approximation,
\( \langle  \sigma_{s,n}^+  \sigma_{s,m}^- \rangle \approx 
 \langle  \sigma_{s,n}^+ \rangle \langle \sigma_{s,m}^- \rangle\)
 for \(n\neq m\). Since the onsite properties of all molecules of the same species are identical,
 we can label \( \langle  \sigma_{s,n}^z  \rangle =  \langle  \sigma_{s}^z  \rangle \),
 and
 \(\langle  \sigma_{s,n}^+ \rangle \langle \sigma_{s,m}^- \rangle = 
 \lvert\langle  \sigma_{s}^+ \rangle\rvert^2\),
 where \(\langle \sigma^{+/z}_s \rangle\) is the expectation for any one of the molecules of type \(s\).
%As \(N_s\to\infty\), we further have
Consider also \(N_s\to\infty\), Eq.~\eqref{eq:PBs1} reduces to
\begin{align}
P_s^B &\approx N_s \lvert\langle  \sigma_{s}^+ \rangle\rvert^2.
\end{align}
To obtain the corresponding dark state population \(P^D_s\), one can simply subtract
\(P^B_s\) from the total molecular population of that species:
\begin{align}
    P^D_s &= P^{\text{tot}}_s - P^B_s \\
    &= \frac{N_s}{2} \left( 1 + \langle  \sigma_{s}^z\rangle \right) -P^B_s \\
    &\approx \frac{N_s}{2} \left( 1 + \langle  \sigma_{s}^z\rangle - 2 \lvert \langle  \sigma_{s}^+ \rangle\rvert^2\right).
\end{align}

\begin{widetext}
    
\begin{figure}
\centering
\hspace*{-.5cm}%
\includegraphics[scale=1]{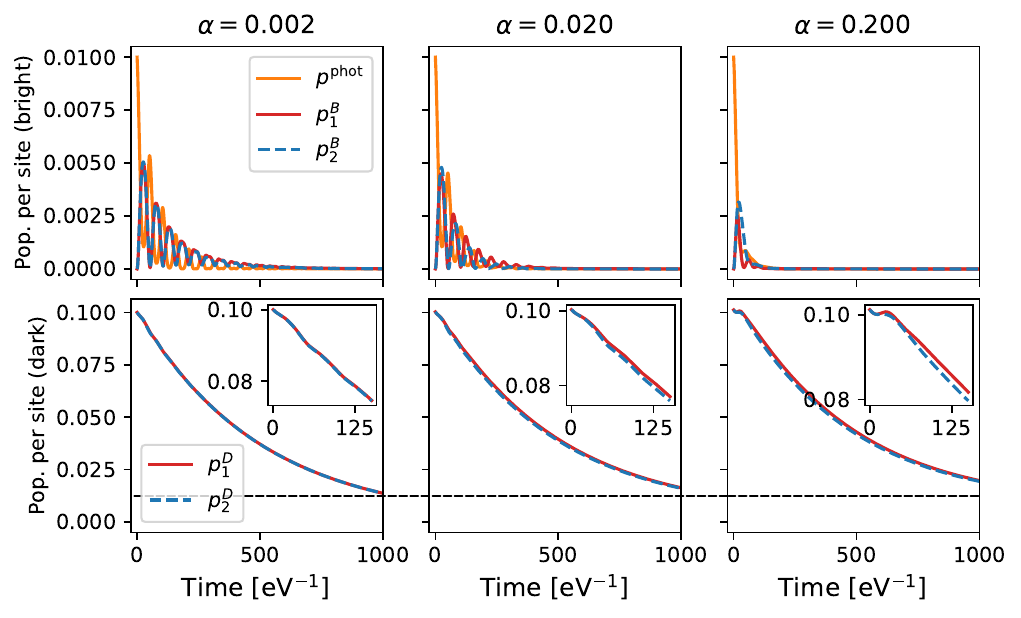}
    \caption{%
    Bright (top row) and dark (bottom row)
    population dynamics in the Markovian (left column), non-Markovian (middle column) and
    large-\(\alpha\) (right column) regimes. The photon mode population is included with the bright 
exciton populations. Insets in each panel of the bottom row highlight small 
oscillations in the dark populations at early times, superimposed on the 
overall exponential decay. Additionally, \(p_1^B\) and \(p_2^B\) feed into \(p_1^D\) 
and \(p_2^D\) at a rate that increases with \(\alpha\). A horizontal dashed line 
across the panels illustrates the enhanced gain of dark states at 
\(\alpha=0.02\) and \(\alpha=0.2\) compared to \(\alpha=0.002\).
    }
    \label{fig:darkbrightpops}
\end{figure}
\end{widetext}

Figure~\ref{fig:darkbrightpops} illustrates the bright and dark populations in the model for energy transfer dynamics across different bath coupling regimes, including Markovian, non-Markovian, and large-\(\alpha\) cases. Our chosen initial conditions set equal populations in the dark excitonic modes, \(p_1^D(0)=p_2^D(0)=0.1\), while the bright excitonic modes start unpopulated. As a result, the dark populations primarily exhibit decay following \(\sim  e^{-\Gamma_\downarrow t}\),
with some gain and small oscillations arising from feeding by, and coherent exchange with, the bright populations.

In the Markovian regime (\(\alpha = 0.002\)), there is minimal feeding of the dark 
states, and the resulting energy transfer (Fig.~\ref{fig:Spectra1}) arises solely 
from the different decay rates of the three polariton modes. In the 
non-Markovian regime (\(\alpha = 0.02\)), feeding of the dark states becomes more 
pronounced, and this loss to the dark modes is linked to the efficient, 
vibrationally induced energy transfer observed in Fig.~\ref{fig:Spectra2}. At 
even larger \(\alpha\), the depletion of bright states occurs so rapidly that 
 coherent oscillations are suppressed, and energy transfer is
 inhibited (Fig.~\ref{fig:Spectra3}). This behavior, reminiscent
 of an overdamped harmonic oscillator, may signal the onset
 of polaron formation.

\end{document}